\documentclass{acm_proc_article-sp}

\usepackage{graphicx}

\begin{document}

\title{The Chemistry Between High School Students and Computer Science}

\numberofauthors{4} 
\author{
\alignauthor
Timothy T. Lenczycki\\
       \affaddr{Poudre High School}\\
       \email{timlen@psdschools.org}
\alignauthor
Kelly Suto\\
       \affaddr{Poudre High School}\\
       \email{kasuto@psdschools.org}
\alignauthor 
Christina Williams\\
       \affaddr{Colorado State University}\\
       \email{williams@cs.colostate.edu}
\and  
\alignauthor Michelle Mills Strout\\
       \affaddr{Colorado State University}\\
       \email{mstrout@cs.colostate.edu}
}


\maketitle
\begin{abstract}
Computer science enrollments have started to rise again,
but the percentage of women undergraduates in computer science is still
low.  Some studies indicate this might be due to a lack of awareness of computer 
science at the high school level.  We present our experiences
running a 5-year, high school outreach program that 
introduces information about computer science within the context of required
chemistry courses.  We developed interactive worksheets using
Molecular Workbench that help the students learn chemistry
and computer science concepts related to  relevant
events such as the gulf oil spill.
Our evaluation of the effectiveness of this approach indicates that
the students do become more aware of computer science as a 
discipline, but system support issues in the classroom can make the
approach difficult for teachers and discouraging for the students.

\end{abstract}




\section{Introduction}

We present a high school outreach program that introduces high school students 
to computer science in the context of interactive worksheets developed
in Molecular Workbench.
High school outreach is important because
the enrollment of underrepresented groups in computer science
 such as women and minorities 
is still quite low. Carter~\cite{Carter06} surveyed 836 high school students 
from a variety of backgrounds
about their impressions of computer science and whether they were considering computer science
as a major.  Unfortunately, 80\% of the students surveyed had ``no idea what
Computer Science majors learned."  
This is a major issue in Colorado due to the lack of computer science courses
being taught in high schools.
From 2005-2011, only 1.5\% of the 116,281 Advanced Placement STEM 
(Science Technology Engineering and Math) exams taken
in Colorado were the computer science exams~\cite{CSAPinfo}.

Based on such observations, we introduced some computer science
concepts to local high school students in coordination with their
current science curriculum. Our team includes two high-school
chemistry teachers, three summer research undergraduates (all women),
and a computer science assistant/associate professor. 
Additionally, during the
summers of 2008 through 2012, we developed interactive worksheets
using Molecular Workbench~\cite{MolWorkBench} and then deployed the
worksheets in the classroom during the school years. The worksheets include 
chemistry and computer science concepts about current events such as the 
gulf oil spill and issues that students find generally interesting such as how mobile 
devices work and what materials are used to make the mobile devices. 
During the summer of 2012, we held a 3-day, hands-on
workshop for a group of 19 junior high and high school teachers to
teach them how to develop interactive worksheets for their
classrooms.

In many regions throughout the country,  related outreach programs are in place or currently in development.  In a three-day summer workshop,
Hart et al.~\cite{Hart08} presented an outreach program where high school math teachers learned ways to incorporate computer science related examples into their courses.  Owens and Matthews~\cite{Owens08}
have developed and piloted a four-week CyberCivics unit for AP 
Government in which high school students explore computer science 
topics and gain hands-on introductory programming experience in the context 
of computing issues related to electronic voting.   To provide support for 
teachers in surrounding districts, they plan to train civics educators to implement 
this unit when they return to their classrooms.   
These examples all share the theme of incorporating computer science 
concepts into current curricular standards.

To evaluate our approach, we developed four different surveys: three
that were given to students over a 5 year period and one that was
administered a month after the summer workshop for the high school
teachers.  The goal of the surveys for the students was to determine
what impact the electronic worksheets with computer science concepts
had on their attitude and understanding of computer science.  The goal
of the survey for the high school teachers was to determine how
feasible such an approach is in the long term.  Are other junior high
and high school teachers willing to use such technology in their
classroom and become computer science ambassadors? We also
qualitatively evaluate the 5 year experience and indicate what worked
and what did not.

\section{Creating the Worksheets}

The goal of the project was to introduce high school students to computer science without 
the benefit of formal computer science courses offered at the high school.
We did this by introducing computer science concepts alongside chemistry concepts
within interactive worksheets.
Some of the worksheets that have resulted from this work 
include\footnote{The URLs provided in footnotes need to be used from within the 
freely available Molecular Workbench software}: 
(1) building atoms\footnote{http://mw2.concord.org/model/1293769e771/page1.cml}, 
(2) evaporation\footnote{http://mw2.concord.org/model/122e684e3d8/gas.cml}, 
(3) movie/entertainment industry\footnote{http://mw2.concord.org/model/13177b6f377/index.cml}, and 
(4) the gulf oil spill\footnote{http://mw2.concord.org/model/13177675544/index.cml}.

\subsection{The Software}
There are numerous interactive animations and simulations available online that utilize drag-and-click manipulation and measurement instruments to provide students with immediate responses to illustrate cause and effect relationships.  Besides MW, two such examples in chemistry are the PhET (Physics Education Technology) Project simulations~\cite{PhET}
 produced through the University of Colorado at Boulder 
and Greenbowe's work at Iowa State University~\cite{Greenbowe}.  These free online resources vary from visually appealing ``real-world" connections to basic interactions between objects within the simulation.  
In addition, the level to which the simulations have been tested prior to posting ranges from extensive instructor and student feedback incorporated into a rating system, to less formal comments posted after teacher use.

We chose the Molecular Workbench software program for building molecular dynamics simulations that model the behavior of submicroscopic particles.  It is designed for teachers to import worksheets ``as is," modify worksheets to fit curricular needs, or create molecular simulations that can then be placed in electronic worksheets developed to illustrate particular scientific concepts.  There is an extensive online tutorial with interactive simulations that supports the development of models and the tools to produce the associated worksheets~\cite{tutorial}.

\subsection{Incorporating Computer Science Concepts}
Using Molecular Workbench software, we developed worksheets each summer for 
use with the students during the school year. One of the goals was to have students 
interact with computational simulation worksheets in tandem with performing physical 
experiments that illustrate the same concepts. The evaporation worksheet follows 
that paradigm.  Simulations were also used directly to convey chemical concepts, 
as in the case of building atoms one sub-atomic particle at a time to teach atomic theory.

The primary goal was to introduce students indirectly to computer science as a 
discipline alongside chemistry concepts using relevant topics.  We developed 
worksheets that described Facebook and internet safety, a worksheet that described 
computer science concepts that are portrayed correctly and incorrectly in movies and 
video games, and a worksheet that described materials used in mobile devices 
alongside computer science concepts such as wireless networking.  Perhaps the best 
example of this was a worksheet that used the gulf oil spill to illustrate how oil and water 
interact.  The worksheet also describes how computer simulations and high performance 
computing are important for predicting the course of the oil spill and where to deploy cleanup crews.

\subsection{Motivating the Teachers}

The chemistry teachers involved in the project consider
teaching chemistry their highest priority.
Since the main goal of this project was to introduce computer
science within the context of a chemistry course, it was important
to keep the priorities of the teachers in mind while designing 
the project.  The teachers found they were motivated to participate
in this project due to a number of factors: (1) the interactive
worksheets seemed to help students with abstract chemistry concepts,
(2) the Colorado curriculum calls for the use of models in K-12 science,
(3) the electronic worksheets helped keep students engaged,
and (4) the high school teachers were able to earn a stipend and course
credit over the summer.

When field tested in Spring 2009, the success of Molecular Workbench in 
chemistry was qualitatively evident.  Students were not only actively engaged 
in the learning process, but made enthusiastic comments to each other with 
phrases like ``Oh, I get it now!" and ``This is fun.  We need to do more of this."  
It seemed that a greater number of students gave correct chemical 
explanations of the relative rates of evaporation in lab reports and appeared 
to struggle less with these sections than in past years on unit assessments.  

One useful aspect of the interactive worksheets was the ability to include hints 
and/or answers so students have the ability to check their own understanding 
of specific concepts and receive immediate feedback at critical checkpoints.
Teachers experience difficulty with larger classes in navigating throughout the 
classroom to answer questions and provide individual feedback, so this enables 
the teacher to monitor for common misconceptions and address these directly 
during class discussion at the conclusion of the activity.

Besides introducing local high school students to computer science as
applied to molecular dynamics simulations, we also wanted to address
specific high school curriculum requirements. 
Standard 5 in the Colorado
K-12 science standards requires that ``models are used to analyze
systems involving change and constancy" for high school
students~\cite{ColoradoStandards}. Using interactive computational
models to help meet this requirement also helps students visualize the
atomic-level interactions that result in the macroscopic-level
observations they make during a laboratory experience.
In chemistry, students have the ability to observe what is happening at the macroscopic level through a laboratory experience but may not make the connection to the abstract, or what is happening at the molecular and atomic level.
MW enables the visualization of microscopic interactions along with the ability of students to interact with these simulations.

The interactive aspect of the models helps keep students engaged,
which is critical due to programs such as the new Response to
Intervention requirement in Poudre School District~\cite{RTI} where
students not performing to their academic potential must be identified
and may require programs and additional staffing that must be
supported by funding not available under current economic conditions.
These additional benefits of this approach to high school outreach
increases the possibility of involving non-computer science faculty at
the high school level.

Finally, everyone's time is valuable.  To maintain a high level of commitment 
for five years, the two high school teacher participants received a 
stipend over the summer and university credit for a group study course. 
The stipend and university credit were planned
and paid for through an NSF CAREER grant.

\subsection{Example Interactive Worksheets}

\subsubsection{Worksheet for Evaporation}

For AP Chemistry, the application of intermolecular forces to evaporation rates was selected as a curriculum target because prior familiarity with existing simulations led to a belief that this particular topic would lend itself easily to molecular modeling.  Previous unit assessments of
student understanding of this topic also indicated this to be an area of conceptual weakness.  The MW activity, which builds on previously developed worksheets~\cite{gaspage}, precedes the actual experiment performed by the students to measure the evaporation rates of different substances.  Utilizing MW, we believe that students will apply an understanding of intermolecular forces to evaporation rates with greater proficiency and to greater conceptual depth than in years past.


Students' fundamental difficulty with understanding intermolecular forces seems to stem from the lack of ability to visualize the particles that interact, as well as the confusion of intermolecular forces with the bonds between the atoms themselves.  By using MW simulations, students interact with visual representations of the forces in meaningful comparisons, which appear to translate into increased concept acquisition and retention through anecdotal evidence and preliminary student feedback.

\subsubsection{Worksheet for Precipitates}
\label{sec:precipitates}

\begin{figure} 
\centering
\includegraphics[width=\columnwidth]{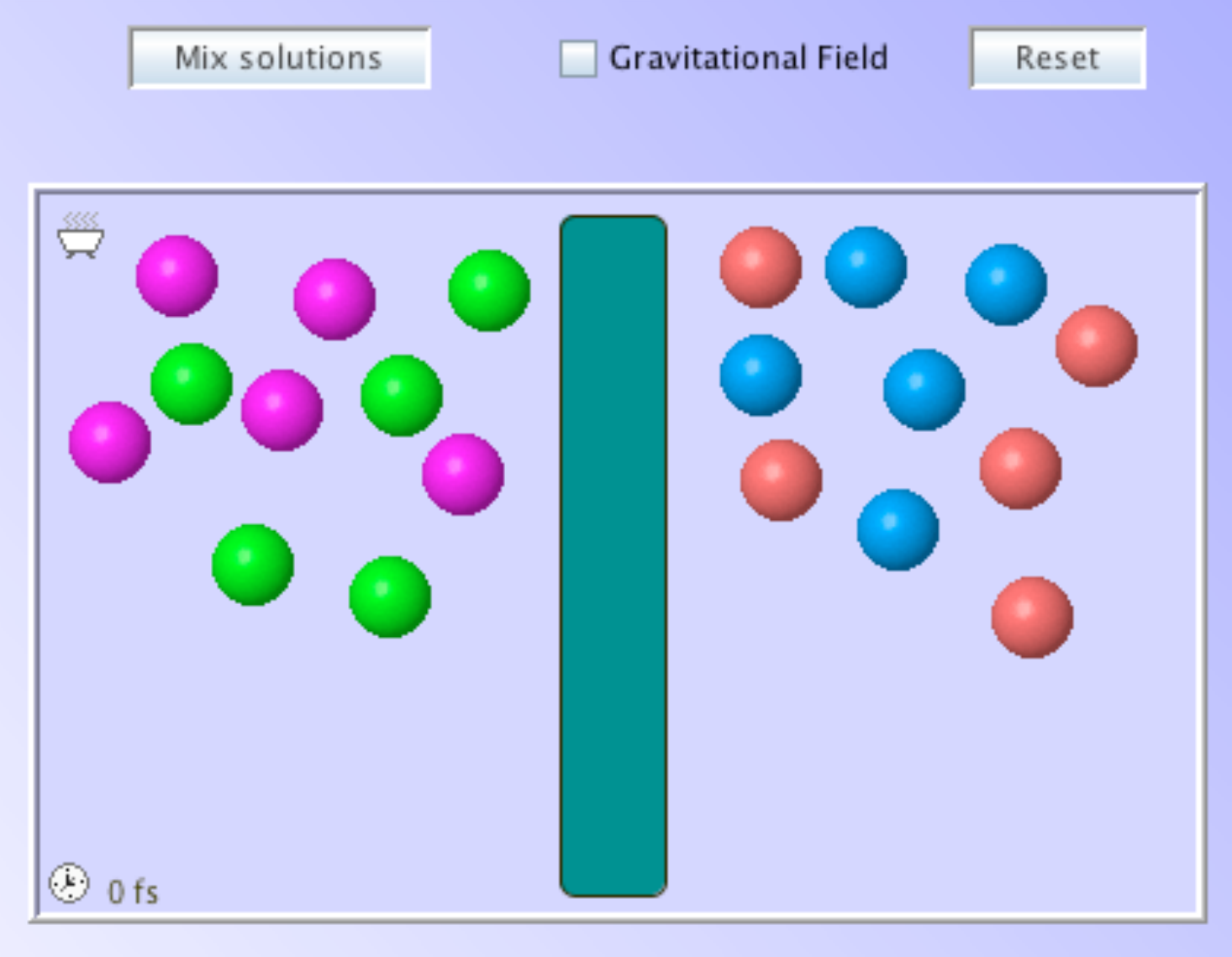} 
\caption{$NaCl$ and $AgNO_{3}$ solutions before they are mixed.} 
\label{fig:before} 
\end{figure}

\begin{figure} 
\centering
\includegraphics[width=\columnwidth]{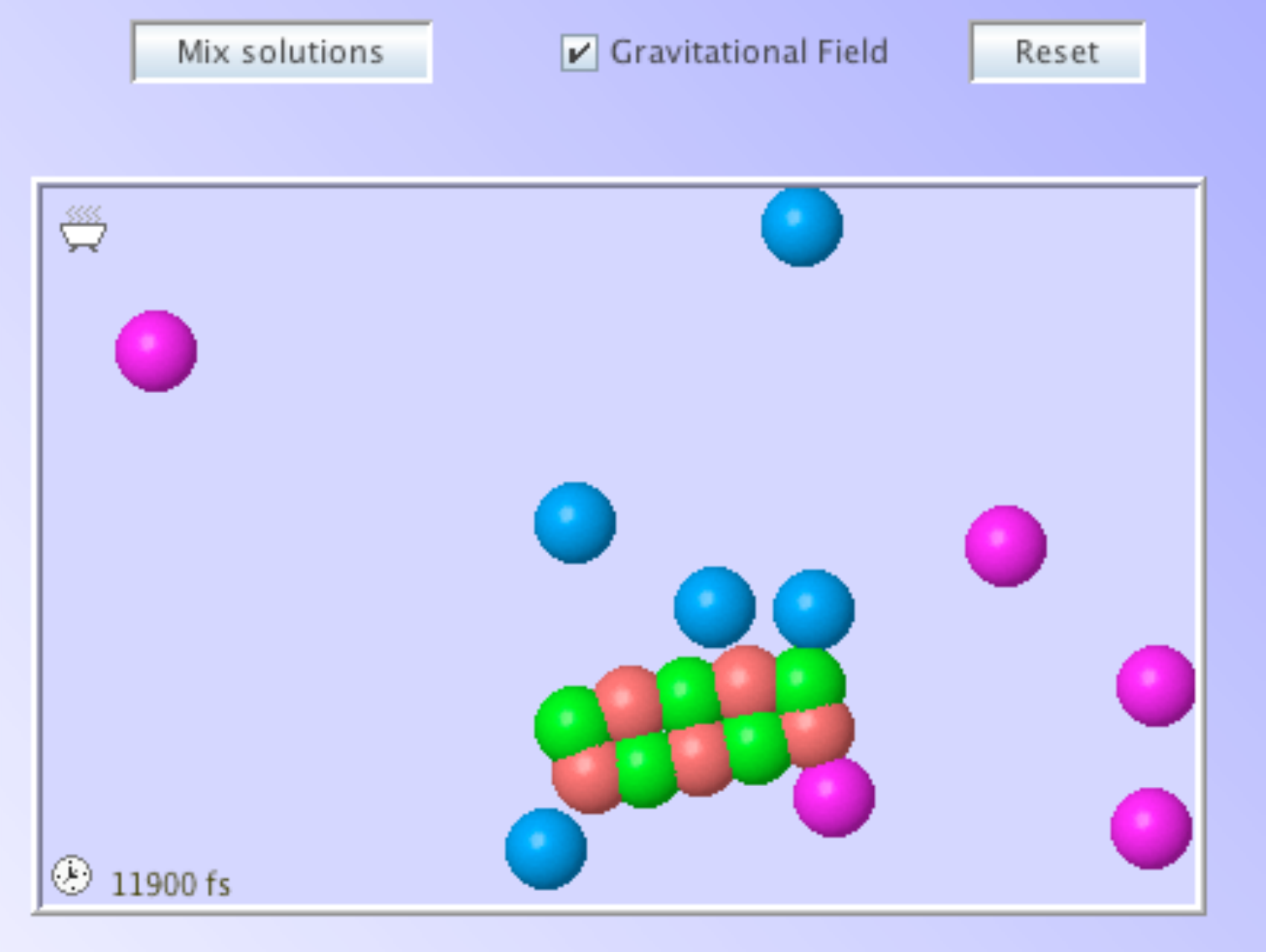} 
\caption{Formation of $AgCl$ precipitate after mixing.}
\label{fig:after} 
\end{figure}

Through the MW worksheet ``Introduction to Solutions," students are led through a series of models and questions related to the solution process to help them understand the properties a particular solute and solvent must have in order for a solute to dissolve.  
Animating the solution process allows the student to have a better grasp of solution properties that differ from that of the solvent alone.  It also represents how chemical reactions occur in solution.  
Figure~\ref{fig:before} shows the dissociation in water of the ionic compound 
silver nitrate, which breaks into $Ag^{+}$ and $NO_{3}^{ -}$.  
Figure~\ref{fig:after} demonstrates the formation of the $AgCl$ precipitate, an insoluble product resulting from the mixing of the previous ionic solutions.  
This worksheet builds off of models previously placed in public domain on the MW website~\cite{watershell, dissolving}. 

During the development of this activity, some limitations of the software became evident.  Because of the small mass of the atoms, gravity has little effect on them.  In order to simulate a precipitate that sinks, we must introduce a larger number of atoms.  At this point, the software no longer runs smoothly and the particles jump from one position to the next.  In order for the molecular simulation engine to achieve the teaching goals of the worksheet, unrealistic parameters needed to be set on both the charges of the ions and the gravitational field strength.  
This was necessary to enable the proper ions ($Ag^{+}$ and $Cl^{-}$) to form the precipitate, the spectator ions ($Na^{+}$ and $NO_{3}^{-}$) to remain in solution, and only the precipitate to fall to the bottom of the container.  
Introducing parallelism could potentially allow the model to operate efficiently with the necessary number of particles and remove the need to set an unrealistic gravitational field.

\subsubsection{Worksheet for Parallelization}

Motivated by the need for more ions in the precipitate worksheet (see Section~\ref{sec:precipitates}) and
the goal of introducing computer science concepts to high school students,
we used the Java concurrent library to parallelize parts of the molecular dynamics simulation engine in MW.
We also added functionality to enable users to initiate the number of threads of parallelism and display its effect on simulation execution time.  For the initial presentation of the parallelization worksheet, we borrowed a quad core from CSU.

The parallelization worksheet first illustrates the ``non-smooth" updates that medium (e.g. 500 atoms) and larger (e.g. 1000 atoms) 
models experience in comparison to a ``smooth" small model (e.g. 100 atoms).
The worksheet then describes how performance problems can be analyzed by profiling the code and analyzing the methods and loops in the computation that require the most execution time.
Finally, the worksheet overviews parallelization and steps the students through an experiment where they modify the number of threads and use the capability of graphing over multiple simulations to observe speedup versus number of threads.

It was not possible to build the parallelization worksheet within the existing MW framework, so we modified the Java implementation to expose control over the parallelization level.  We also created a new graph type within MW to plot results such as speedup over multiple runs of a simulation. 
In our modified version it is now possible for a user to
specify the number of threads to use when running simulations. The built-in capabilities of the scripting
language allow parallelism to be set from various easy-to-use places such as sliders, spinners, etc.
The second part of the support for parallelism is the ability to graph the effects parallelism has on simulations.
Previously, the Molecular Workbench had a built in graphing capability; however, it only was designed to
graph data that changes as a simulation runs (such as potential energy) rather than data that changes
between simulations. 
We added a new form of graph; this one collects data for each simulation at the end of a time
limit specified by the user, and displays the data in real-time. It can be graphed as a scatter plot or a mean
of all data for each number of threads.

\section{Sharing What We Learned}

For the culmination of the five-year outreach project, a three-day summer
workshop was offered to high
school science teachers within Poudre School District.
The workshop was held June 18-20, 2012 in the Computer Science building at
Colorado State University. 
A total of 18 teachers participated in addition to the original
two high school teachers that ran the workshop.

The original two high school teachers wanted to share their
experiences in the classroom using Molecular Workbench especially
in light of the school district's push for more technology in the classroom.
With the
approval of a 2010 Mill Levy and Bond issue, the Poudre School
District has committed to ``refresh technology and provide technology
support for student learning opportunities" and ``purchase and install
technology in district schools to provide learning opportunities for
students."\footnote{http://www.psdschools.org/about-us/board-leadership/major-
initiatives/2010-mill-levy-bond}  
One focus of the district is to
provide a laptop computer for every high school student, starting with
the incoming freshmen of the 2011-2012 school year. This commitment
requires educators to investigate technology-based teaching
methodologies. 
The idea for the three-day workshop was to
 to provide support and time for teachers who will be
incorporating these technologies into their curriculum this fall. 

On the first day, we
told the teachers about the overall outreach project and about
computer science.  Teachers were encouraged to suggest computer
science as a possible discipline for students.
Then we 
introduced teachers to the Molecular Workbench application and the
idea of the electronic worksheet. Teachers were encouraged to explore
Molecular Workbench simulations and worksheets in addition to other
online simulations. Teachers spent the next day modifying existing
electronic worksheets or developing paper and pencil worksheets and
electronic worksheets to use in conjunction with existing online
simulations. 

During the final day of the workshop teachers shared the
lessons produced, in addition to their concerns about and difficulties
associated with technology-based lessons. Teachers were able to
utilize the time to produce lessons geared to a variety of skill
levels and science content. For example, a physics teacher developed
pen and paper worksheets to be used in conjunction with PhET
simulations. Two ninth grade biology teachers modified an existing
Molecular Workbench electronic worksheet on cellular transport
mechanisms. Another teacher developed an electronic worksheet to be
used with photosynthesis and respiration simulations accessed through
another website. These lessons were then placed in a Dropbox folder to
provide common access to lessons generated through the workshop.

Teacher comments during the workshop included, ``Besides learning about
some great learning tools, and learning from other great educators, it
was just really fun." and ``Best workshop in a long time!  Appreciate
it!" 
The teachers as a group requested that this specific workshop be offered
each summer.  The small stipend, single university course credit, and
lunch each day was a nice silver lining and showed that we respected the
teachers time and effort.

A survey taken a month after the workshop found that all of the
survey participants (7 teachers)
 felt the workshop had the right
 balance of instruction and worktime.
 Six people felt the workshop was just the right length, and one
 person felt it could be longer.  Everyone was willing or
 very willing to attend a similar workshop in the future.
 There were mixed responses in terms of whether the credit
 and/or stipends were needed.  The university credit seemed to
 be the most attractive and important incentive.
 Most importantly, everyone indicated that they would be using
 the worksheets in their classrooms at least once per year,
 with 2 people indicating they would be using one each week.

\section{Evaluating the Impact on Computer Science Attitudes}

The main goal of the project was to introduce high school students
to computer science within the context of their required chemistry courses.
We evaluated the impact of this approach using three surveys.
The first survey was developed 
by an undergraduate research assistant.
The second survey was developed by another undergraduate summer
student, but was based on the
work of Hoegh and Moskal at the Colorado School of Mines~\cite{Hoegh09}, 
where
they developed a survey tool to measure the attitudes of high school 
students about computer science.
The third survey was developed after doing a statistical analysis of
the results of the second survey.

In this section, we describe the results of each of the surveys.  Unfortunately,
the approach we developed did not have as much of an impact 
on computer science attitudes as
we had hoped.  
Though this approach introduced the students to computer science 
and cleared up some misconceptions about the field, technical difficulties 
experienced with the machines and the network used in the high school 
to deploy the interactive worksheets caused some negative results in 
terms of how students viewed the usefulness of computer science.
 
\subsection{The 2008-2009 Survey}
During the spring semester of 2009, the computer science associate professor
and an undergraduate research assistant visited Poudre High School and 
had students in AP Chemistry and one IB MYP (the Middle Years
Program of the International Baccalaureate Program) Chemistry section complete 
a survey based on what the students knew about computer science and its 
applications.  Two example questions were, ``When you think of
a computer scientist, what best describes what comes to your mind?
[Someone who mostly writes computer programs, Someone who mostly does
research, Someone who mostly installs/fixes computers or computer systems, 
Someone who mostly builds computers, Someone who mostly helps people when their 
computers aren't working, I don't know, Other"] 
and
``Have you ever thought about being a computer scientist for a career? 
If so, are you planning to pursue this career path?".
Out of the 43 students who took the survey, 18 thought of a computer scientist as
someone who mostly writes computer programs and 12 thought of a computer scientist as
someone who does research.  About 7 of the 43 students indicated that they were interested
in becoming a computer scientist.

In
the first session, the students worked with the evaporation worksheet, and in the second session, they worked with the precipitates and parallelization worksheets.  The students were also asked about what they liked and did not like in the worksheet as well as for suggestions to improve the worksheet.  
Many of the students liked the graphical/visual interface and the interactivity. 
The dislikes were more varied and included the desire for more interactivity, for a bug with the thermometer to be fixed, they wanted harder questions, and they did not like that they had to print out the finished worksheet\footnote{MW does have online submission, and we used that in later sessions at the high school.}.
Their suggestions included making more interactive activities, providing more facts/explanation, figure out technical issues\footnote{Some of the computers had internet connectivity issues.}, and present more real-life applications.
The second group of students liked the worksheets because they were interesting, included simulations, and had diagrams.  They disliked the worksheets because they had to read a lot, they ``didn't like Molecular Workbench," they ``did not finish so didn't see the point of it," and ``felt that the longer questions were difficult to understand."

\subsection{The 2010-2011 Survey}

To more accurately measure the impact of this program on high school students'
attitudes toward computer science, we developed
a survey based on the
work of Hoegh and Moskal at the Colorado School of Mines~\cite{Hoegh09}.  Ideally,
nineteen questions were to be asked both before and after Molecular
Workbench activities to gauge student attitudes toward computer
science.  During the 2010-2011 school year, only pre-activity data
was reliably gathered and analyzed to determine the validity of the
survey instrument itself.

This second survey included the following quote: ``Computer science is the 
study of computation -- what can be computed and how to
compute it."~\cite{Wing06}
%
The quote was included to give the students some idea of what computer science is since
most of these students had never taken a computer science course.

Table~\ref{table:2010} shows the survey, which
 was designed to measure three constructs
of student attitudes toward computer science: understanding, interest,
and usefulness.  The {\em understanding} construct included four questions
(4, 8, 11, and 15), {\em interest} comprised six questions (3, 5, 7, 10,
12, and 14), and {\em usefulness} included four questions (2, 6, 9, and
13).  The remaining questions gathered demographic information or
prompted students to begin thinking about computer science
applications and thus were not grouped into a particular construct. 
Items themselves were not actually questions but rather statements
whereby students were asked to respond using a four point Likert
scale: strongly disagree, disagree, agree, and strongly agree.
 A neutral category was not included in order to encourage students to
make a positive or negative decision.  Many of the questions were
adapted from the work of Hoegh and Moskal.  

\begin{table}[t]
\caption{Questions on the 2010-2011 survey.}
\label{table:2010}
\begin{tabular}{|c|p{.85\columnwidth}|}

\hline
\textbf{\#} & \textbf{Question}     \\ 
\hline
1 & Question 1 did not use the Likert scale for its answers. \\
\hline 
2 & Developing computing skills will be important to my career goals. \\
\hline 
3 & I hope I can find a career that provides the opportunity for me to use computer
science concepts.\\
\hline
4 & I am comfortable learning computing concepts such as setting up a home network
and developing an excel spreadsheet with macros. \\
\hline
5 & Students who are skilled at computer science are just as popular as other
students.\\
\hline
6 & I use my computer for solving day-to-day problems at home/school. \\
\hline
7 & I am interested in learning computer science concepts and skills. \\
\hline
8 & I find it easy to use computer applications for class assignments. \\
\hline
9 & Computer science concepts and skills are useful for everyone in every profession
today. \\
\hline
10 & I would like to see how computer science concepts are used in the development
of video games and movies. \\
\hline
11 & I understand that performing long division and other arithmetic tasks by hand
can be described with a procedure, or algorithm, which in turn can be programmed
into a computer. \\
\hline
12 & I am likely to take a computer science class in college. \\
\hline
13 & Developing computer science skills will help me secure a better job. \\
\hline
14 & I find that using computers within the context of other courses helps aid
understanding of concepts not directly related to computing such as chemistry,
history, and english. \\
\hline
15 & I am comfortable learning computer science concepts and skills. \\
\hline

\end{tabular}
\end{table}


The responses to the questions were coded on a scale from one to four,
with the highest score reflecting the most positive response.  The
sample size was 150 students, grades ten through twelve, male and
female, taken from General Chemistry, IB MYP Chemistry, and Advanced
Placement Chemistry.  




As was done in the Hoegh and Moskal~\cite{Hoegh09} paper, we analyzed the Cronbach's 
alpha for each set of questions to determine how well that question fit within 
each construct.  The usefulness construct was seen as problematic since the 
responses to those four questions together do not meet the 0.7 threshold. It is 
interesting to note that the entire data set taken together had the highest 
Cronbach's alpha, suggesting that the entire sample can be viewed as mapping to a single construct.

Upon doing a factor analysis to determine which questions
correlate, we found that instead of having three constructs, we appeared
to have only two constructs.  Therefore we decided to only pull two
constructs from this initial survey  
for a more rigorously administered pre and post survey.  Table~\ref{table:2011}
shows the questions that were selected for the 2011-2012 school year
survey and the reasons why the questions were selected in parentheses.

\begin{table}[th!]
\caption{Questions on the 2011-2012 survey.  The factor loadings are from factor analysis
and > 0.4 is considered significant.}
\label{table:2011}
\begin{tabular}{|c|p{.85\columnwidth}|}

\hline
\textbf{\#} & \textbf{Question}     \\ 
\hline 

& Interest \\
\hline
\hline
 1 & I am interested in learning computer science concepts and skills.  (Question 7 in Table~\ref{table:2010})\\
\hline
4 & I hope I can find a career that provides the opportunity for me to use computer science concepts.  (Question 3 in Table~\ref{table:2010}) \\
\hline
 7 & I am likely to take a computer science class in college.  (Loaded very well on the Interest Factor in the post-survey data, 0.837) \\
\hline
 10 & Developing computing skills will be important to my career goals.  (Question 2  in Table~\ref{table:2010}, even though it was previously grouped in the usefulness construct) \\
\hline
 13 & I like to use computer science to solve problems.  (from Heersink and Moskal) \\
\hline
\hline
& Understanding \\
\hline
\hline
2 & I find it easy to use computer applications for class assignments.  (Question 8 in Table~\ref{table:2010}) \\
\hline
 5 & I use a computer for solving day-to-day problems at home/school.  (Question 6 in Table~\ref{table:2010}, even though it was grouped in the usefulness construct.  The word ``my" is changed to ``a" to avoid discrimination between those with and without computers.)\\
\hline
 8 & I am comfortable learning computer science concepts.  (Question 4 in Table~\ref{table:2010}) \\
\hline
11 & Developing computer science skills will help me secure a better job.  (Usefulness question in the post-survey that loaded very well on the Understanding Factor, 0.708) \\
\hline
 14 & I use computer science skills in my daily life.  (adapted from Heershink and Moskal) \\
\hline
\hline
& Misconceptions \\
\hline
\hline
3 & Computer science is primarily about writing programs.\\
\hline
6 & Computer scientists spend all their time sitting in front of a computer. \\
\hline
9 & Most computer scientists have poor social skills. \\
\hline
12 & Most computer science jobs are being outsourced to other countries. \\
\hline
15 & Computer professionals are not likely to be involved in solving real world problems. \\
\hline
\end{tabular}
\end{table}

We decided to use  a 15
question survey encompassing 3 constructs: Interest, Understanding,
and Misconceptions.  
We decided to give the survey one
week prior to the delivery of the Molecular Workbench lesson, and the
identical post-survey within one week after the lesson.  

\subsection{The 2011-2012 Survey}

Based on the survey analysis from the initial trial, an improved
survey was given to students during the 2011-2012 academic year to
assess the change in interest in, understanding of, and misconceptions
around Computer Science through using computer simulations to learn
Chemistry concepts with Molecular Workbench (MW) software.  Students
were given identical surveys both before and after an hour long
interactive series of MW worksheets ranging in content from atomic
structure, intermolecular forces, and nuclear chemistry for separate
populations. 
Ideally, the pre-survey was to be given within one week
prior to the MW experience and the post-survey within one week after. 
Problems with the hardware, both access to machines and network
reliability, prevented ideal protocol from being followed in two of
the three student populations.

The goal with the 2011-2012 survey was to determine if the
electronic worksheets had an impact on student attitudes about 
computer science.  To measure this we used a paired t-test, where
we looked at the pre and post survey data for each student and
 tried to determine if the overall  results were significant.
 A low p-value, <0.05, is statistically significant.

The entire data set represented three distinct sections: pre and post
survey data from General Chemistry students during an atomic structure
unit, pre and post survey data from MYP Chemistry students during a
nuclear chemistry unit, and pre and post survey data from AP students
during a unit on intermolecular forces.  Individual student responses
were correlated and the paired T test was performed on the entire data
set representing 101 separate students.  (Approximately one-third of
the data represented only pre-survey information and was removed.) 
Only questions 1 and 3 showed significant change in response (p-value
< 0.05), with mean delta values of -0.238 and -0.178 respectively. 
This corresponded to a decrease in Computer Science interest as
measured by question 1 and a decreased belief that Computer Science is
only about writing programs as measured by question 3.  The decreased
interest was somewhat reinforced by a marginally significant negative
result when the entire interest construct (questions 1, 4, 7, and 10)
was analyzed with the paired T test: p-value = 0.0525, mean $\Delta$ =
-0.0644.

When the data was split into its three distinct student populations,
additional results could be gleaned.  Some data was determined to span
outside the time periods represented by these three populations and
was thus removed.  Using only the MYP Chemistry data (N = 26), the
paired T test showed marginally significant results (p-value = 0.0501
in both cases) for questions 1 and 12.  Interest again decreased (mean
$\Delta$ = -0.269) and the misconception response measured in question 12
increased by the same amount (mean $\Delta$ = 0.269).  No significant result
could be gathered when the constructs were analyzed as a whole.  It
must be noted that a time lag of two months between pre and post
survey data most certainly impacted the MYP Chemistry results.  The
cause of this time lag was the lack of availability of dependable
computers on which to perform the surveys.

Significant results (p-value < 0.05) were seen with only the General
Chemistry students for questions 1, 8, 14, and the interest construct
as a whole (N = 17).  Interest in computer science went down as
measured by question 1 and the entire construct, mean $\Delta$ values =
-0.529 and -0.206 respectively.  If computer science understanding is
indeed measured by question 8 (it did not load on the appropriate
factor), it went up by mean $\Delta$ = 0.235.  This, however, was balanced by
a decrease in the average change in understanding responses as
measured by question 14, mean $\Delta$ = -0.235.  This is confirmed by an
insignificant change in the understanding construct as a whole
(p-value = 0.255).  Lack of availability of dependable computers also
impacted the General Chemistry results.  The small number of machines
available meant that the pre and post surveys were performed in teams,
with each response representing a combination of two student opinions.
 The reliability of these results were therefore also questioned.

Finally, when only the AP student data (N = 29) was analyzed,
questions 7, 12, 13, 14, and the understanding construct (questions 2,
5, and 14 combined) each showed significant change.  Here ideal survey
protocol was observed.  
The interest in Computer Science seemed to increase slightly, though
question 13 did not load uniquely on the interest factor.  Belief in
the misconception described in question 12 decreased, while
understanding of Computer Science increased as measured by question 14
and the entire construct.

Within the second-generation 15 question student Computer Science
attitude survey at the high school level, 7 questions emerged as
viable measures of what we labeled as Interest in Computer Science and
Understanding of Computer Science as a field.  Cronbach's alpha
measurements >0.7 supported the reliability of questions within each
construct.  Factor analysis >0.4 confirmed the loading of each
question on its appropriate factor.

When the data was separated into its three distinct student
populations, the paired T test on each population revealed little if
any significant change in student attitudes toward Computer Science
when having a Molecular Workbench computer simulation experience in a
Chemistry setting.  In the most reliable data set, the AP Chemistry
students, what we labeled as understanding of the field of Computer
Science was shown to have increased slightly, corresponding to a
significant decrease in a particular Computer Science misconception. 
A perceived increase in student interest in Computer Science as
measured on questions 7 and 13 among the AP population is balanced by
the negative finding on question 1 among the entire data set
encompassing all the Chemistry students surveyed.  It is important to
note that the lack of availability of dependable computers and network
access at the high school level not only prevents the collection of
reliable data but may be a contributing factor in the interest of
students in the field of Computer Science.

%
%

\section{Conclusions}

	Computational models are critical to the way science is currently done in university and industry research labs, but their use in K-12 classrooms has been limited.  Molecular Workbench provides an excellent platform for integrating computational models and particle simulations into the high school classroom.  Electronic worksheets are designed to teach students scientific concepts through the interactive manipulation of models and simulations, either reinforcing conventional teaching techniques or replacing them altogether.  Research shows that concept acquisition is much more effective through active rather than passive student engagement~\cite{learn}.  

The use of computational models at the high school level provides a vector for 
increasing awareness of the applications of computer science and computer science 
concepts.  Student survey results (in a school where there is no computer science course) 
show that there is a varied understanding of what computer science entails and 
interest in computer science.  Student surveys indicate that some computer science
misconceptions were fixed throughout this process, but the barrier of getting the
technology to reliably work is a larger issue than expected.

\section{Acknowledgments}
This work is sponsored by NSF CAREER grant
CCF 0746693.
We would like to thank Shinjini Kar and Kate Ericson for their
aid in putting together worksheets and survey questions for this project.
We would also like to thank Christopher Wilcox for his work parallelizing MW.

%

\end{document}